# Spin Magnetotransport in a Two-Dimensional Electron System under quantum Hall regime and with Rashba spin-orbit effect.

M. A. Hidalgo[(1)] and R. Cangas[(2)]

[(1)] *Departamento de Física, Universidad de Alcalá, Alcalá de Henares (Madrid), Spain.*
[(2)] *Departamento de Física, Escuela Universitaria de Ingeniería Técnica Industrial, Universidad Politécnica de Madrid, Spain.*

Electronic mail: miguel.hidalgo@uah.es, roberto.cangas@upm.es

**Abstract**

The article shows a simple model that computes the magnetoconduction and the integer quantum Hall effect (IQHE) in a two-dimensional electron system (2DES) where the spin is other degree of freedom in the system. The 2DES is confined in a quantum well (QW) immersed in a heterostructure, where the Rashba spin-orbit interaction is present. When an external magnetic field is applied to the system, the competition between the spin-orbit interaction and the Zeeman effect on the magnetoconduction of the 2DES is analysed. Also the model reproduces the case where two sub-bands are occupied in the QW. In this case different spin oriented 2DES concerned to each subband can be treated independently, assuming the whole system as the sum of four independent systems (two subbands plus two 2DES spin systems in each subband). The model has been tested with experimental results obtained from a 2DES formed in an InGaAs layer for one and two subbands.

## 1. Introduction

The advances in the fabrication of mesoscopic systems with few impurities and defects leads to the macroscopic observation of microscopic quantum effects, such as the quantum Hall effect (QHE) and Subnikov-de Haas (SdH) oscillations [1]. The quantum Hall effect is one of the most amazing and interesting phenomena in the condensed matter physics discovered at the end of the past century. The integer quantum Hall effect (IQHE) is characterized by the appearance of quantized plateaux in multiples of $\nu e^2/h$ in the non-diagonal magnetoconductivity $\sigma_{xy}$ ($h$ is the Planck constant, $e$ the electron charge and $\nu = 1,2,3,...$ an integer), and vanishing values in the diagonal magnetoconductivity $\sigma_{xx}$ observed at the same magnetic field ranges, [1]. The magnetoconductivities are measured on a two-dimensional electron system (2DES) at very low temperatures (quantum Hall regime). These phenomena are related to the magnetoconductance of charged particles in a two dimensional electron system (2DES), where the spin of the charge carriers play a relevant role, and is the responsible of the appearance of even and odd plateaux in the IQHE. It is also possible to manipulate not only the charge current in the devices, but also the spin of the carriers by means of magnetic and/or electric fields. In fact, in 1990 Datta and Datta [2] proposed a spin-polarized field effect transistor (FET). The gate electrode on the top of the FET device is used to control, by means of an electric field, the spin of the electrons. This electric field induces a spin-orbit interaction (SOI) that breaks the spin degeneration of the energy states in the 2DES. Even without any external magnetic/electric field, the carriers of the 2DES are also spin polarized by the internal built-in electric field due the structure inversion asymmetry (SIA) of the semiconductor heterostructure. The first theoretical study of this effect was made by Rashba [3] in 1960 (the SOI due to SIA is called Rashba effect). In 1989 Das et al. obtained an evidence of spin splitting carrier populations at zero magnetic fields in InGaAs/InAlAs heterostructures [4]. The SIA electric field is normal to the 2DES confined in the inversion layer or quantum well, and the spin splitting provided by this field is given by the expression [5]:

$$\Delta E_{so}^{SIA} = 2\alpha k \qquad (1)$$

where $\alpha$ is a parameter which depends on the electric field asymmetry of the heterostructure, and $\vec{k} = (k_x, k_y)$ is the 2DES wave vector, $\alpha$ is also called Rashba parameter. Measured values of $\alpha$ varies between $2\times10^{-12}$ eVm and $5\times10^{-11}$ eVm for a 2DES confined in InGaAs/InAlAs heterostructures [6,7]. Also in MOSFET devices $\alpha$ can be tuned with the gate voltage [2,8]. On the other hand, zinc-blend semiconductors have bulk inversion asymmetry (BIA). Due this asymmetry the local electric field varies along the crystal directions and therefore the SOI (Dresselhaus effect) [9].
For a 2DES confined in a $(x,y)$ plane the energy is spin-orbit splitted by this local field, giving $\Delta E_{so}^{BIA} = \gamma(k_x k_y)k$ [10], where $\gamma$ is a parameter that depends on the material. The BIA effect is stronger than the SIA effect in the GaAs/AlGaAs heterostructure 2DES [11], and the values measured of spin-split energy in this alloy are of the order of $20\,\mu eV$ at the Fermi level [12]. In quantum well and heterostructure devices made with InGaAs/InAlAs systems the SIA effect has more relevance than BIA, obtaining spin-split energies of the order of *meV* at Fermi level [5]. This has hindered spintronics research in the GaAs/AlGaAs system, which provides the highest mobility 2DES.

More recent devices are proposed by Schliemann et al. [13] and Nitta et al. [14], both based in the spin manipulation by means of an electric field. Schielemann has proposed a spin-field-effect transistor based on SOI of both SIA and BIA types, where the spin-independent scattering processes have not influence in the spin transport, and also shows how the interplay between SIA and BIA can lead to *k*-independent spin wave functions. Nitta has proposed a device based in the interference of spinning currents guided in narrow wires rings.

The present work analyses the electrical magnetoconductance (magnetoresistance) behaviour of a 2DES confined in a heterostructure quantum well (QW), under QHE conditions, and with Rashba SOI effect (at low temperature) using a simple model based on semiclassical considerations and taking into account the spin orientation degree of freedom. The model reproduces the SDH oscillations experimental data obtained in devices where the 2DES is confined in a QW where one or two subbands are filled by electrons. Also the model shows the Hall magnetoresistance when one or two subbands are filled in the QW.

From the theoretical point of view several attempts to understand SdH oscillations of the magnetoconductance and the IQHE have been published. The most accepted is based on the 'gendanken' experiment thought up by Laughlin [15], where the 2DES localized states due to ionized impurities and defects play a crucial role to explain the plateaux of the Hall magetoconductivity (magnetoresistivity) and the SdH oscillations of the diagonal magnetoconductivity (magnetoresistivity) with minima values close to zero. However, experimental evidences show that the measures made on 2DES with higher electron mobility (materials with few defects and impurities) provides better plateaux precision. The model that we proposed does not use localized states to explain the QHE and SdH effects, but a simple one-electron theory with two assumptions: first, the existence of a flow of carriers from/to the QW to/from the heterostructure where it is immersed (the heterostructure behaves as a "source/drain" of charges), and where long relative variations in the 2DES carrier concentration occurs with negligible variations in the 3D carriers density of the environment, keeping constant the chemical potential which is fixed by the enviroment [16] (*M. A. Hidalgo, Microelectronic Engineering 43-44 (1998) 453*); second, external magnetic fields and/or SOI lifts the spin degeneration, splitting the 2DES in two independent 2DES, one with parallel spin and the other with antiparallel one [17-19]. The first assumption indicates a constant value of Fermi level in the 2DES when the 2D carrier concentration changes when the applied magnetic field varies. The second assumption leads to consider the electron system as the sum of two 2DES independent spin sub-systems.

With this model we reproduce the experimental results of SdH oscillations obtained by Nitta et al. [14] measured in a device where the 2DES is confined in a QW where only is filled by electrons the first energy level (subband); and also reproduce the experimental data obtained by Can Min Hu et al. [20] where there are two filled subbands in the well that confines the 2DES. The model also reproduces the integer quantum Hall magnetoresistivity.

### 2. Magnetotransport (one subband is filled in the QW)

At zero external magnetic fields we consider the 2DES confined in a QW like a two dimensional non interacting electron gas under the effective mass approximation, perturbed by impurities, defects and the spin-orbit coupling. Then, the energy of each electron can be approach as $E(k) = \hbar^2 k^2 / 2m + U \pm \Delta E_{SO}/2$, where $m$ is the electron effective mass, $U$ takes into account the electrostatic interaction with impurities and defects, and $\Delta E_{SO}$ is the spin-orbit split energy caused by the SIA effect. Hence, each energy level is split in two levels spaced in energy by a factor $\Delta E_{SO}$. Then the whole 2DES can be studied like two 2DES, one for each spin orientation (parallel and anti-parallel to the spin-orbit magnetic field). On the other hand, when an external magnetic field $B$ is applied normal to the 2DES, and assuming no spin-orbit coupling effects, the energy of the system is discretized in Landau Levels (LL), with values $E_{N_{LL}}^{\pm} = (N_{LL} + 1/2)\hbar\omega \pm 1/2 g\mu B$, where the last term (Zeeman term) correspond to spin ↑↓ orientations, $N_{LL} = 0,1,2,3...$, $g$ is the effective g-factor, $\omega = eB/m$ the cyclotron frequency and $\mu$ the Bohr magneton. Measured values of $g$-factor varies from -0.44 [21] in GaAs to -30 in InAs alloys [22], and depends on the carrier concentration [23]. In a 2DES confined in a heterostructure device, taking into account the SIA and Zeeman effects, the energy of carriers is obtained by the expression [5] (*Bychkov Y. A. et al., J. Phys. C 17 (1984) 6039*).

$$E_{N_L}^s = \hbar\omega \left[ N_L + s\frac{1}{2}\sqrt{\left(1 - |g|\frac{m}{2m_0}\right)^2 + \frac{\gamma}{B}N_L} \right] \quad (2)$$

with $s = \pm 1$ for $N_L = 1,2,3,....$, $s = +1$ for $N_L = 0$, $\gamma = 8\alpha^2 m^2/\hbar^3 e$ and $m_0$ is the rest electron mass. Figure 1 shows a fan of energy levels of the 2DES with data obtained from Table 1 for $V_G = 0.3\ V$ (see below) [14] and where the 2DES is confined in the In$_{0.53}$Ga$_{0.47}$As layer of the In$_{0.53}$Ga$_{0.47}$As/In$_{0.52}$Al$_{0.48}$As heterostructure at 0.4 K. A regular Hall bar sample was made with this structure. An applied voltage $V_G$ on top of device induces a variation on the Rashba parameter $\alpha$ and on the carrier concentration $n$. In the figure a crossing of spin-up ($E_{N_L}^+$) and spin-down ($E_{N_L'}^-$, $N_L \neq N_L'$) energy levels is appreciated for some values of the magnetic field.

It is well known that the density of states (DOS) in the 2DES at zero magnetic field (and without Rashba effect) is $D_0 = m/2\pi\hbar^2$ i.e. the states are uniformly distributed in energies (in this equation we have not taken into account the spin degeneration). When only exists Rashba effect, the DOS is [24]:

$$D_\pm(E) = \frac{1}{4\pi\mu_c}\left[1 \mp \frac{\alpha}{\sqrt{4\mu_c E + \alpha^2}}\right], \quad E \geq 0$$

$$D_-(E) = \frac{1}{2\pi\mu_c}\frac{\alpha}{\sqrt{4\mu_c E + \alpha^2}}, \quad E < 0 \quad (3)$$

The DOS converges to the constant value $D_0$ (no spin degeneration is considered) when $\alpha$ is zero. But when a magnetic field is applied and the SIA effect is taking into account the energy states are given by Eq. (2). In this approximation we assume that each level is degenerated in $eB/h$ [25]. The DOS of a 2DES under the application of a magnetic field normal to the system has a shape like a "comb", where the pinned "teeth" are related to the $E_{N_L}^s$ values, and can be modelized with an "*ad hoc*" gaussian shape function [23]:

$$D(E)_s = \frac{eB}{h}\sum_s \sum_{N_L} \left[\frac{\pi}{2}\Gamma_{N_L s}^2\right]^{-1/2} \exp\left\{-2\frac{(E - E_{N_L}^s)^2}{\Gamma_{N_L s}^2}\right\} \quad (4)$$

where $\Gamma_{N_L s}$ is the width of the $E_{N_L}^s$ level. The level broadening is strongly dependent on the range of scattering potentials. For short range scatters ($d < l/\sqrt{(2N_L + 1)}$ where $d$ is the order of the range and $l = \sqrt{\hbar/eB}$ the magnetic length) $\Gamma_{N_L s}^2$ depends on the strength of the magnetic field. The broadening due to long range potentials is proportional to fluctuations of the local potential energy $((V(r)-\langle V(r)\rangle)^2$, and can be considered negligible in $\delta$-doped samples where the impurities are far from the 2DES. Then, we use the expression

$$\Gamma_{N_L s} = \Gamma_0 + \kappa \sqrt{(2\hbar^2/\pi)(\omega/\tau)} \qquad (5)$$

Where $\Gamma_0$ and $\kappa$ are fitting parameters, and $\tau$ is the relaxation time. Eq. (5) involves two 2DES with two different spin states. Figures 2a and 2b show the density of states of a 2DES for the data given in Table 1 for a gate voltage $V_G = 0.3\,V$ at 1.2 T and 1.8 T, respectively.

TABLE 1

| $V_G(volts)$ | $n(m^{-2})$ | $\alpha(eVm)$ | $\Gamma_{N_L s}$ |
|---|---|---|---|
| 0.3 | $2.0 \times 10^{16}$ | $7.2 \times 10^{-12}$ | $0.010 E_F + \sqrt{(2\hbar^2/\pi)(\omega/\tau)}$ |
| 0.0 | $1.9 \times 10^{16}$ | $7.7 \times 10^{-12}$ | $0.014 E_F + 1.4\sqrt{(2\hbar^2/\pi)(\omega/\tau)}$ |
| -0.3 | $1.8 \times 10^{16}$ | $8.3 \times 10^{-12}$ | $0.016 E_F + 1.6\sqrt{(2\hbar^2/\pi)(\omega/\tau)}$ |

Table 1. Parameters used to compute the magnetoresistance of the 2DES confined in a In$_{0.53}$Ga$_{0.47}$As layer of the device described in reference [14]. The effective mass of the carriers is $m = 0.05 m_0$, the effective $g$-factor is -4. The relaxation time is $\tau = 1.0 \times 10^{-12} s$, obtained from the measures made by Burgt et al. [26].

When the applied magnetic field increases, the energy levels $E_{N_L}^s$ showed in the DOS move to the Fermi level ($E_F$), and the conduction occurs when each level crosses $E_F$, providing a modulated oscillation in the magnetoconductivity (beating pattern of SdH oscillations, see Figure 3). The maxima (minima) values of magnetoconduction occur when there are coincidence of spin-up and spin downs levels at Fermi level, i.e. when $E_{N_L}^+ = E_{N'_L}^- \approx E_F$ ($N_L \neq N'_L$). This occurs in a region of magnetic field near to $1.2 - 1.3$ T, related to values $N_L, N'_L \approx 31-34$, and corresponds to the DOS showed in Fig. 2a. The beating pattern arises from the existence of two kind of carriers, and the sum of their concentrations at Fermi level. The nodes of the magnetoconduction oscillations occur in the region of magnetic field in which there is no coincidence of the DOS energy levels at Fermi energy, i.e. when $E_{N_L}^+ \neq E_{N'_L}^-$. This condition can be write $(E_{N_L}^+ + E_{N_L-1}^+)/2 = E_{N_{LnC}}^- \approx E_F$, where $N_{LnC} = N_L + 1, N_L + 2,...$ The first node occurs close to 1.8 T, when the levels $N_L \approx 22-23$ cross the Fermi level and $N_{LnC} = N_L + 1$, as can be seen in Fig. 2b. The second node occurs at 0.95 T with $N_L \approx 43$ and $N_{LnC} = N_L + 2$. High order nodes occurs for $N_{LnC} \geq N_L + 3$ and $B < 0.95\,T$.

Also exists a competition between Rashba and Zeeman effects which occurs by the coincidence of levels $E^+_{N_L}$ and $E^-_{N_L+1}$. If we compare Eq. (2) with the conventional spin-up and spin-down energy states associated with $N_L$ LL number, this correspond to $E^+_{N_L}$ and $E^-_{N_L+1}$ states, i.e. $\Delta E_{spin} = |E^+_{N_L} - E^-_{N_L+1}|$ [24]. In the absence of Rashba effect, Eq. (2) reproduces well known LL energy spectrum. In the limit of large magnetic fields the Zeeman term dominates the spin splitting, obtaining $\Delta E_{spin} = g\mu B$. In the opposite limit, when $B \to 0$, $\Delta E_{spin} = 2\alpha k_F$ is obtained, where $k_F = \sqrt{2\pi n}$ is the Fermi wave vector and "$n$" is the equilibrium carrier concentration. The condition of coincidence is $E^+_{N_L} = E^-_{N_L+1}$, and is governed by the equation:

$$\sqrt{\left(1 - |g|\frac{m}{2m_0}\right)^2 + \frac{\gamma}{B}N_L} + \sqrt{\left(1 - |g|\frac{m}{2m_0}\right)^2 + \frac{\gamma}{B}(N_L + 1)} = 2 \qquad (6)$$

With the data of Table 1 for $V_G = 0.3\ V$, this occurs at 4.85 T and $N_L \approx 7-8$.

The magnetoconductivities are obtained relating the carrier current density with the applied electric and magnetic fields. The general expression is :

$$\vec{j}(\vec{r},t) = \vec{j}_\uparrow + \vec{j}_\downarrow = e\sum_s \int_{-\infty}^{\infty} \vec{v}_s f(E) D(E)_s dE \qquad (7)$$

where $\uparrow\downarrow$ are related to the spin orientation, $\vec{v}$ is the carrier velocity, $E$ is the electron energy and $f$ the distribution function perturbed by the electric and magnetic fields [25]. Hence, we are assuming two currents with different spin (parallel and antiparallel to magnetic field). To compute the magnetoconductivity we use the semiclassical theory, using the Eq. (4) as density of states. Taking into account the linear relationship $\vec{j} = [\sigma]\vec{E}$, where $E$ is the applied electric field and $[\sigma]$ the magnetoconductivity tensor, we obtain:

$$\sigma_{xx} = \sigma_{yy} = \frac{e^2 N\tau}{m}\frac{1}{1+(\omega\tau)^2} \qquad (8a)$$

$$\sigma_{xy} = -\sigma_{yx} = \frac{e^2 n\tau}{m}\frac{\omega\tau}{1+(\omega\tau)^2} \qquad (8b)$$

$n$ is the whole equilibrium carrier concentration and $N$ the carried concentration at the Fermi level, given by the expressions:

$$n = n_+ + n_- = \sum_s \int_{-\infty}^{\infty} f_0(E) D(E)_s dE \qquad (9a)$$

$$N = N_+ + N_- = \sum_s \int_{-\infty}^{\infty} D(E)_s E\left(-\frac{\partial f_0}{\partial E}\right) dE \qquad (9b)$$

The magnetoresistivity tensor is obtained by the relationship $[\rho]=[\sigma]^{-1}$, obtaining

$$\rho_{xx} = \rho_{yy} = \frac{\sigma_{xx}}{\sigma_{xx}^2 + \sigma_{xy}^2} \tag{10a}$$

$$\rho_{xy} = -\rho_{yx} = -\frac{\sigma_{xy}}{\sigma_{xx}^2 + \sigma_{xy}^2} \tag{10b}$$

Figure 3 shows the computed values of magnetoresistivity obtained from Eq. 10a of a 2DES confined in the In$_{0.53}$Ga$_{0.47}$As heterostructure described in reference [14]. From this computation we determine the data shown in Table 1 for three gate voltages, obtaining a good agreement with the experimental data. In fact, the model also reproduces the experimental results of the variation of the Rashba parameter with the applied gate voltage.

At $V_G = 0.3$ the maximun of local oscillations occurs at a value of magnetic field near to 1.4 T instead in the range of 1.2-1.3 T as we expected if the coincidence condition $E_{N_L}^+ = E_{N_L+2}^- = E_F$ is applied. This is a consequence of the overlapping of adjoining levels and the increasing of the density of states as the magnetic field grows. On the other hand, as we have seen before, the Rashba-Zeeman competition at Fermi levels occurs near to 4.85 T. Shen et al. [26] deduced the appearance of a resonance in the spin Hall conductance at values of magnetic field where exist the coincidence $E_{N_L}^+ = E_{N_L+1}^-$, being Eq. (6) the resonant condition at Fermi level. The effect that we deduced is an increase on the magnetoconductivity at this value of the field, observed in clean samples. Figure 4 shows a plot of $\sigma_{xx}$ computed using a gaussian width of a tenth of $\Gamma_{N_{Ls}}$.

The theoretical model also modelizes the integer quantum Hall effect (IQHE). Figure 5 shows the Hall magnetoresistivity in a 2DES with a carrier concentration of $2.0 \times 10^{16} \, m^{-2}$, obtained at three different values of Rashba parameter. In order to resolve even and odd plateaux (even and odd $\nu$, respectively) we use in Fig. 5 small gaussian width of energy levels in the DOS, and a g-factor equal to 8. When $\alpha = 0$ the spin degeneration is broken only by the Zeeman effect, and the width of the plateaux grows with the magnetic field. The odd plateaux became wider when B grows. When $\alpha \neq 0$ the width of the plateaux varies due to the effect of the Rashba spin-orbit, producing the disappearance of the odd plateaux in the magnetic field regions where the Rashba-Zeeman competition occurs, i.e., when $E_{N_L}^+ = E_{N_L+1}^-$ near to Fermi level. For $\alpha = 3.0 \times 10^{-11} eVm$ the Rashba-Zeeman competition occurs near to 14.5 T, vanishing the $\nu = 5$ plateaux, and for $\alpha = 5.0 \times 10^{-11} eVm$ occurs at values close to 25 T, vanishing the $\nu = 3$ plateaux. As expected, the value of the

Hall magnetoresistivity is not affected by the SIA spin-orbit effect ($\rho_{xy} = h/(v\,e^2) = 25812.807/v$ Ω, $v = 1, 2, 3, ...$), but the width of the plateaux change when α varies. In fact, the quantum Hall effect is reproduced in dirty samples where local high electric fields due to impurities exist.

Figure 6 shows the effect of the impurities and defects in the Hall magnetoresistivity. This effect is introduced in the model by means of the relaxation time and hence the width of the energy levels in the DOS. Figure 6 shows the Hall magnetoresistivity of the 2DES system without Rashba effect ($\alpha = 0$) using three relaxation times.

## 3. Magnetotransport (two or more subbands are filled in the QW).

If the 2DES is confined in a QW with subbands energy levels $E_i$ ($i = 1, 2,...$), the eigenvalues of (1), assuming only the Rashba effect, are given by the expression

$$E_{iN_L}^s = E_i + \hbar\omega \left[ N_L + s\frac{1}{2}\sqrt{\left(1 - |g^*|\frac{m^*}{2m_0}\right)^2 + \frac{\gamma}{B}N_L} \right] \qquad (11)$$

In a QW with two filled subbands with energies $E_1$ and $E_2$, the 2DES can be considered as the sum of four 2DES, which are related to the $E_{1\uparrow}$, $E_{1\downarrow}$, $E_{2\uparrow}$ and $E_{2\downarrow}$ states. Hence the whole DOS (at zero magnetic field) of the four subsystems is computed by the expression:

$$D(E) = \sum_s \sum_i D_{is}(E) \qquad (12)$$

Fig. 7 shows the DOS of the 2DES confined in a QW with two subbands (with energies $E_1$ and $E_2$). The whole density of states is considered as the sum of the four independent DOS related to the four 2DES.

In the present case DOS of a 2DES under the application of a magnetic field can be considered as the sum of the 2DOS spin independent for each subband.

$$D(E)_{is} = \left(\frac{eB}{h}\right) \sum_{s} \sum_{N_L} \left[\frac{\pi}{2}\Gamma_{N_L s}^2\right]^{-1/2} \exp\left\{-2\frac{(E-E_{iN_L}^s)^2}{\Gamma_{N_L s}^2}\right\} \qquad (13)$$

Figures 8a to 8f show the evolution of the DOS presented in Fig.7 when an external magnetic field is applied and exists Rashba effect. To model the DOS we have used a Rashba parameter of $\alpha = 0.7 \times 10^{-11} eVm$ and a effective g-factor $g^* = 4$. In order to compute the width of the gaussian function of energy levels, we use the fitting parameters $\Gamma_0 = 0.01 E_F$, $\kappa = 1$. The relaxation time is again $\tau_{is} = 10^{-12} s$ and we will assume the same for all the subbands.

Figures 8a and 8b show the oscillations and nodes of the DOS. The maxima and minima values of the oscillations occur when the energy levels of the different spin coincide, i.e. $D(E)$ a maximum value when $E = E_{N_L}^{\pm} = E_{N'_L}^{\mp}$, $(N_L \neq N'_L)$, in each subband, and also when there are coincidence in energy of the maxima values of the DOS in the two subbands, and at the same time occur the coincidence of the minima values in the oscillations. The nodes occur when there is not a coincidence of energy states in the DOS, i.e, when $E = E_{N_L}^{\pm} \neq E_{N'_L}^{\mp}$. The number of nodes and their positions depend on the energy balance between Rashba and Zeeman terms. Rashba term grows with the momentum $k$, and hence with the kinetic energy, while Zeeman term keeps constant.

Figures 8c and 8d show with clearness the energy levels in both subbands, in Fig. 8c there is an overlapping of the $E_{iN_L}^+$ and $E_{iN_L}^-$ levels in the DOS in each subband and when both subbands are added. In Fig. 8d there is coincidence of energy levels of different spin in each subband, i.e., $E_{iN_L}^{\pm} = E_{iN'_L}^{\mp}$, $(N_L \neq N'_L)$, but there are not overlapping of the levels of the two subbands. Figures 8e and 8f show the DOS at high magnetic fields (8 T and 12 T respectively). The height of the DOS levels depends again on the coincidence of levels intrasubband and the overlapping of levels intersubbands. As we will see below this DOS behaviour and its value at Fermi level explains the magnetoconductance of the 2DES.

In order to obtain the magnetoconductivity of the 2DES formed in the semiconductor heterostructure, we have to calculate the density of carriers. Assuming that the 2DES is confined in a QW with two filled subbands, it can be considered each subband energy level as a pocket that contains two "independent" 2DES, with spins parallel and antiparallel to the magnetic field. Therefore, the whole carrier concentration confined in the QW is given by the sum of the four 2DES concentrations:

$$n = \sum_s \sum_i n_{is} \tag{14}$$

On the other hand the total carrier concentration $N$ at Fermi level $E_F$ is given by the expression :

$$N = \sum_s \sum_i N_{is} \tag{15}$$

If the carrier concentration at zero external magnetic field is known, the Fermi level of the system is determined from Eq. (14). To compute the magnetoresistance we use experimental data obtained by Can Min Hu et al. [20]. The 2DES is formed in a 20 nm thickness $In_{0.53}Ga_{0.47}As$ layer where the two subbands levels are filled. The whole electron concentration at zero magnetic field is $n_0 = 3.6 \times 10^{16} m^{-2}$ and the carrier concentration of the subbands are $n_1 = 2.8 \times 10^{16} m^{-2}$ and $n_2 = 8 \times 10^{15} m^{-2}$. The calculated Fermi level is $E_F = 0.172$ eV and the computed subband levels are $E_1 = 0.038$ eV and $E_2 = 0.134$ eV. The effective mass is $0.05 m_0$. Figure 9a shows the evolution of the total carrier concentration ($n$) in the whole 2DES when the external magnetic field increases, the evolution of the the two subbands carrier concentrations ($n_1$ and $n_2$), and the evolution of the spin up/down 2DES that forms each subband ($n_{1+}, n_{1-}, n_{2+}, n_{2-}$).

Figure 9b shows the evolution of the carrier concentration computed at Fermi Level ($N$) in the whole 2DES, in the two subbands ($N_1$ and $N_2$), and the evolution of the spin up/down 2DES that forms each subband ($N_{1+}, N_{1-}, N_{2+}, N_{2-}$). The values of $N_1$ and $N$ show a beating pattern with a node near to 2.2 T. The nodes occur when there are no coincidence of the levels $E^+_{N_L}$ and $E^-_{N'_L}$, at Fermi level. The $N$ value also shows an envelope modulation created by the sum of the $N_2$ value.

The magnetoconductivity tensor is computed by

$$[\sigma] = \sum_s \sum_i [\sigma]_{i,s} \tag{16}$$

and the components of the tensor are:

$$\sigma_{xx} = \sigma_{yy} = \sum_s \sum_i \frac{e^2 N_{is} \tau_{is}}{m^*(1 + (\omega \tau_{is})^2)} \tag{17a}$$

$$\sigma_{xy} = -\sigma_{yx} = \sum_s \sum_i \frac{e^2 n_{is} \tau_{is} \omega \tau_{is}}{m^*(1 + (\omega \tau_{is})^2)} \tag{17b}$$

The current density for the two subband problem can be expressed:

$$\mathbf{j} = [\sigma] \mathbf{E} = (\mathbf{j}_1 + \mathbf{j}_2)_\uparrow + (\mathbf{j}_1 + \mathbf{j}_2)_\downarrow \tag{18}$$

The magnetoresistivities are obtained by the relationship between tensors $[\rho] = [\sigma]^{-1}$.

We reproduce the value of the magnetoresistivity given in reference [20] when the whole 2DES carrier concentration is $3.6 \times 10^{16} m^{-2}$ and two energy subbands in the QW are filled (we have assumed the same relaxation time for each subband) .Figure 10a shows the SdH oscillations of the magnetoresisitivity with a visible node near to 2.2 T, and figure 10b shows a detailed plot in the interval 0.6 T – 1.5 T of the magnetic field, where it can be appreciated two more nodes at values near to 0.75 T and 1.1 T respectively. As we mentioned before, the nodes occur when there is no coincidence between energy levels at $E_F$, i.e. when $E_{N_L}^+ \neq E_{N_L}^-$ at Fermi level. The appearance and definition of the nodes depends on the overlapping and the width $\Gamma_{N_L s}$ of the DOS energy levels.

Figure 11a shows the calculated Hall magnetoconductivity ($\sigma_{xy}$) of the whole 2DES and the Hall magnetoconductivities ($\sigma_{xy1}, \sigma_{xy2}$) related to the two subbands (obtained with the experimental data [20] used above to compute the SdH oscillations). Each $\sigma_{xy1}$ and $\sigma_{xy2}$ behaviour corresponds to the pattern of the integer quantum Hall effect, with plateaux that have values of magnetoconductivity equal to $\nu e^2/h$, $\nu = 1, 2, 3, \ldots$. This is the result that we expect because each 2DES is treated independently, and the model reproduces the integer QHE in the case of only one filled subband. On the other hand, the total Hall magnetoconductivity $\sigma_{xy}$ is the sum $\sigma_{xy1}$ and $\sigma_{xy2}$ and also has plateaux with values $\nu e^2/h$, although with less resolution. Figure 11b shows the Hall magnetoresistivity of the 2DES, where it can be appreciated plateaux with values of magnetoresistivity equal to $h/(\nu e^2)$. In both figures we have selected the interval of magnetic field from 5 T to 25 T to have well resolved plateaux at low filling factors $\nu$.

## 4. Conclusions

In conclusion, we have developed a simple semiclassical theory that reproduces the magnetoconduction of a 2DES confined in a QW when one or two subbands are occupied and when the competition between Rashba and Zeeman effects are both significant. Then in the model that we use the spin plays an important role in the magnetoconduction. The model starts with the whole carrier concentration at zero external magnetic field, that establish the Fermi level. When two subbands are occupied, the carrier concentration of each subband is obtained from the value of the subband energy level respect to Fermi level. Each subband is considered as the sum of two independent 2DES with different spin polarizations due to Rahsba effect. Therefore we consider the whole 2DES confined in a QW as two filled subbands, and hence four independent 2DES. The evolution of the DOS with the external applied magnetic field explains the SdH oscillations and the integer QHE. As it is shown along the paper the model is able to reproduce accurately experimental data. And, eventually, our model can be generalized to systems with more than two filled subbands.

**FIGURES WITH CAPTIONS**

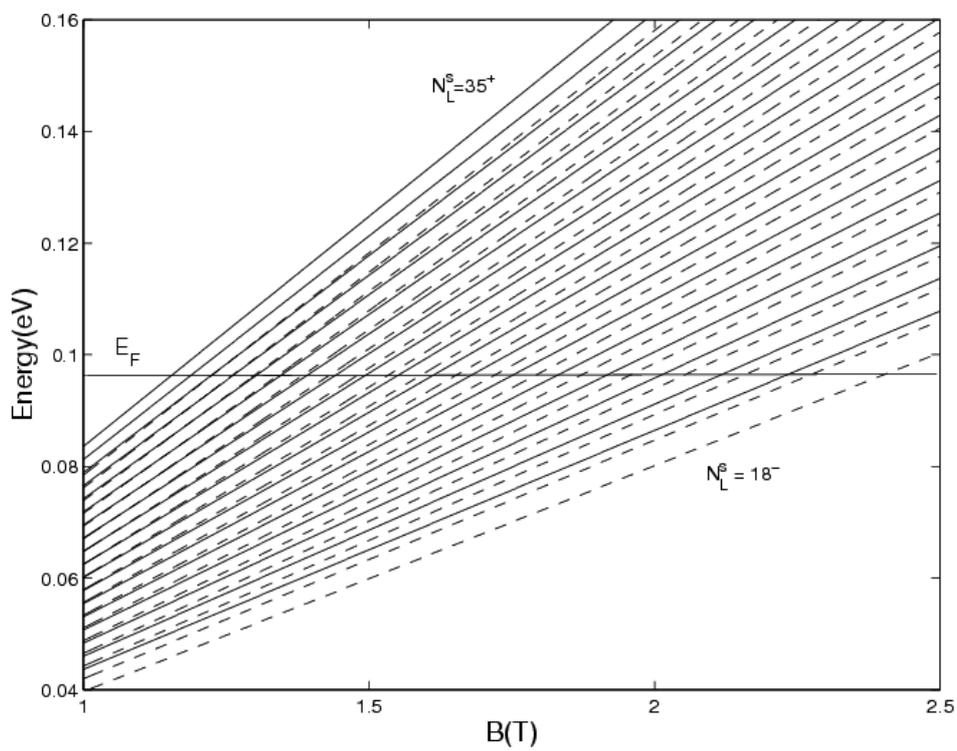

**Figure 1**. Plot of the energy levels of the 2DES system with the data of Table 1 for $V_G = 0.3\ V$. Dash line corresponds to $E^-_{N_L}$ levels and solid line to $E^+_{N_L}$ levels. It shows the crossing of spin-up and spin-down levels when $B$ decreases.

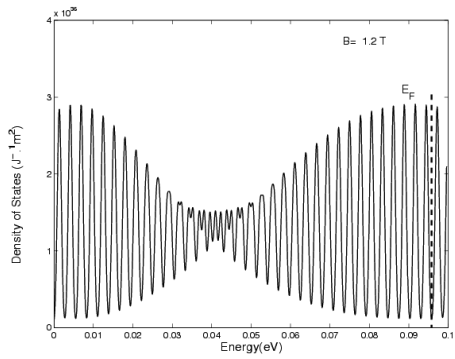 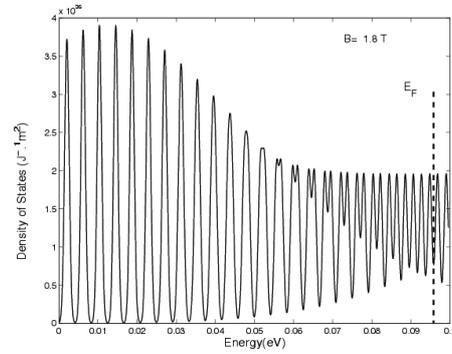

Figure 2a

Figure 2b

**Figure 2**. 2DES density of states with data from Table 1 for $V_G = 0.3$ V.
Figure 2a, at $B = 1.2\,T$ where there is coincidence of adjoining levels at energies close to $E_F$. Figure 2b, at $B = 1.8\,T$, where all levels are spin resolved at energies near $E_F$ and hence there is no coincidence of levels.

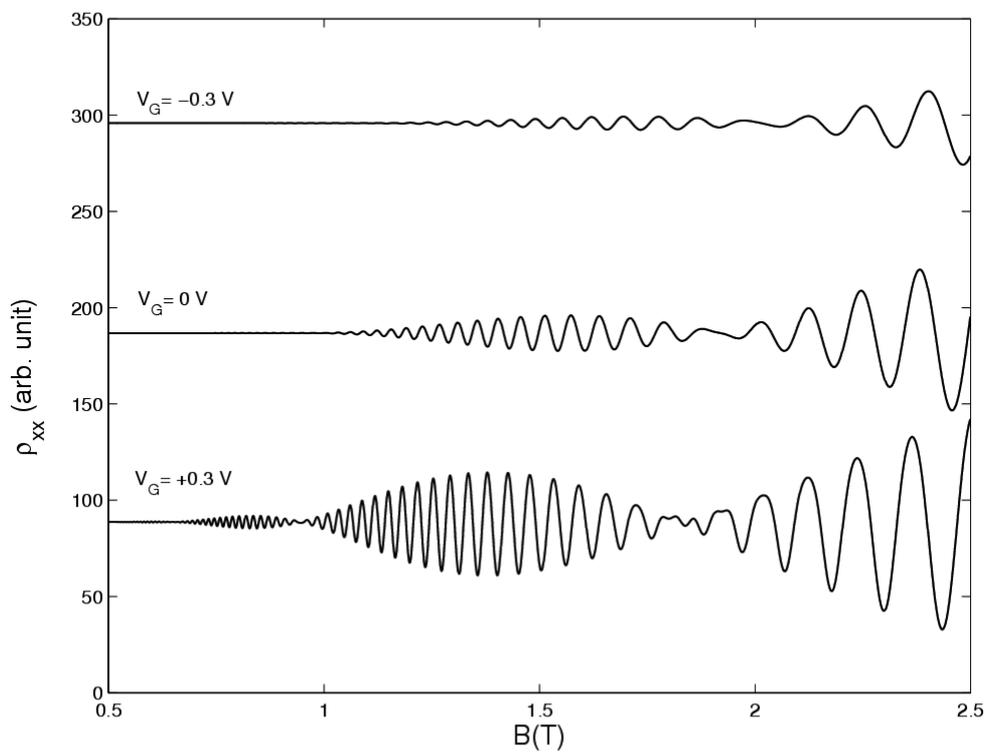

**Figure 3.** SdH oscillations of a 2DES confined in the $In_{0.53}Ga_{0.47}As$ layer of a heterostructure computed with data obtained in reference [7] and using the parameters given in Table 1. This model reproduces the experimental values in which a variation of Rashba parameter is observed with the gate voltage $V_G$ applied to the system.

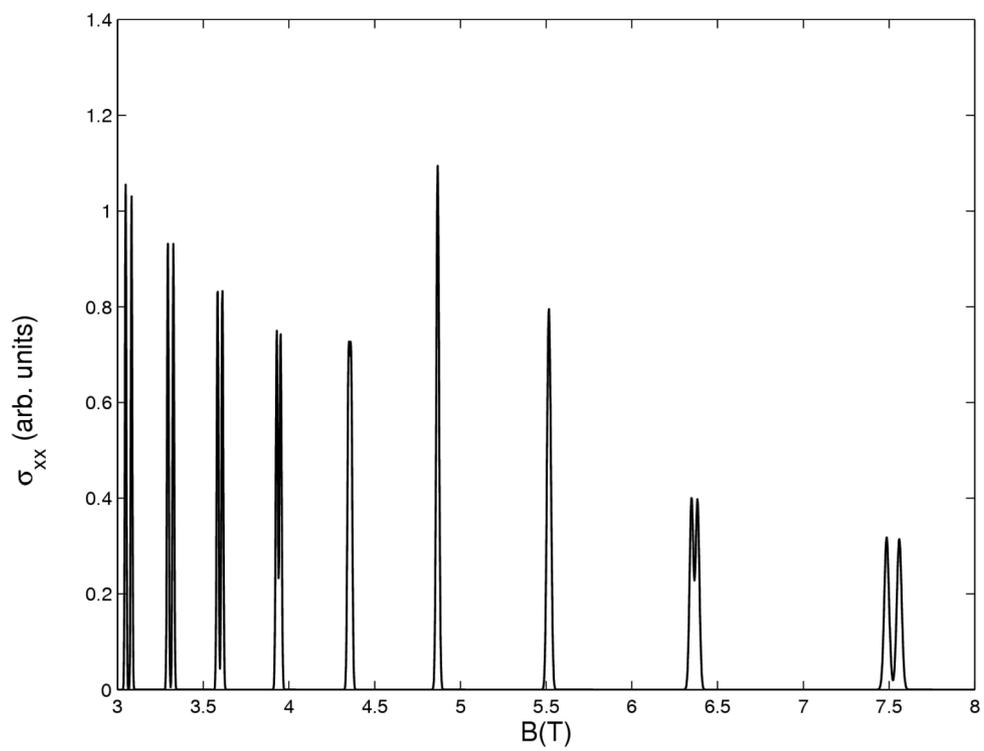

**Figure 4** Computed magnetoconductivity for a clean 2DES. The higher value of the magnetoconductivity near 4.85 T is due to the competition of Rashba and Zeeman effects. (The data used correspond to $V_G = 0.3\ V$ and $0.1\Gamma_{N_{Ls}}$ from Table 1.).

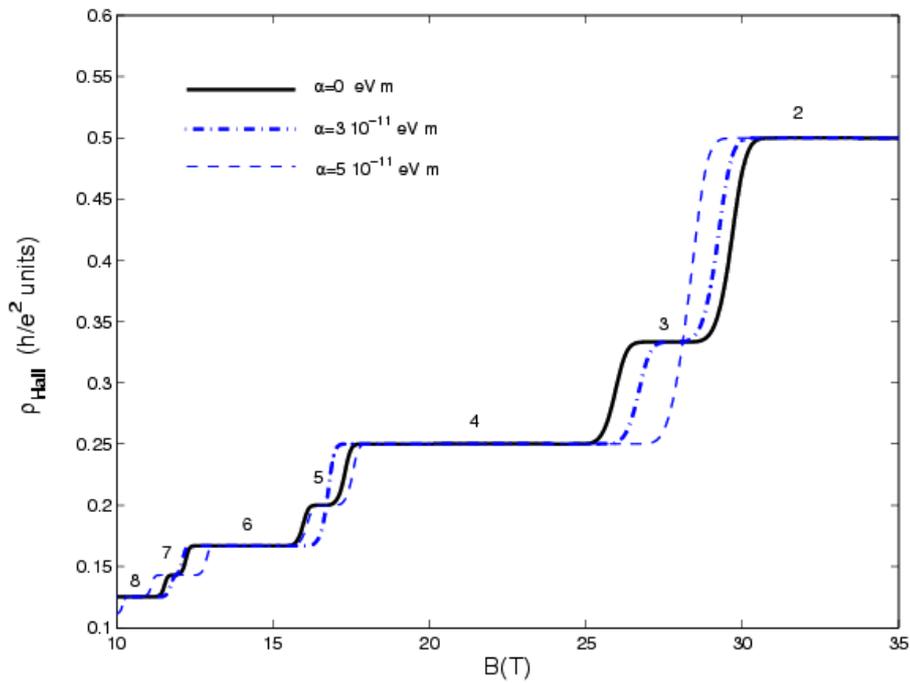

**Figure 5** Quantum Hall magnetoresistivity at three different Rashba parameters and an effective *g*-factor of -8. Continuous line correspond to $\alpha = 0$. For $\alpha = 3\times 10^{-11} eVm$ (dash-dot line) and $\alpha = 5\times 10^{-11} eVm$ (dot line) the $\nu = 5$ and $\nu = 3$ plateaux vanishes, respectively. This occurs in both cases due to the Rashba-Zeeman competition at these values of the magnetic field.

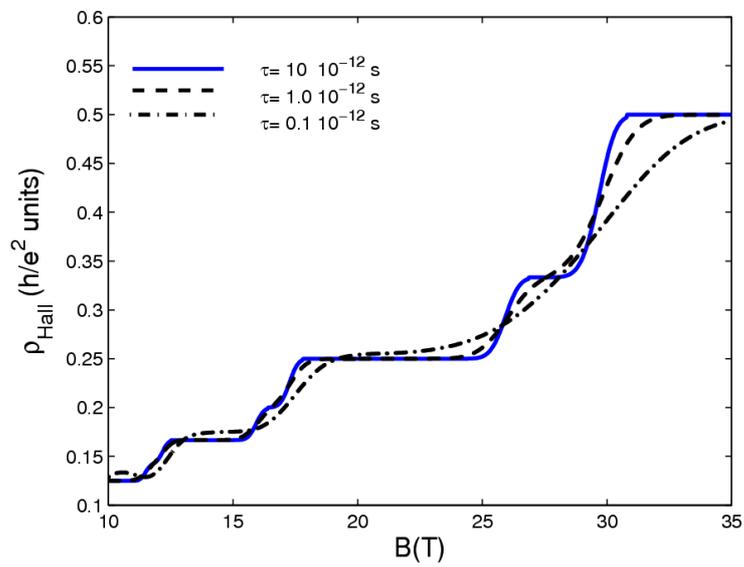

**Figure 6.** Hall magnetoresistance computed at three relaxations times $\tau$, with values of $10\times10^{-12}\,s$ (cotinuous line), $1.0\times10^{-12}\,s$ (dash line) and $0{,}1\times10^{-12}\,s$ (dash dot line). The width of the energy levels is computed by $0.02 E_F + \sqrt{(2\hbar^2/\pi)(\omega/\tau)}$.

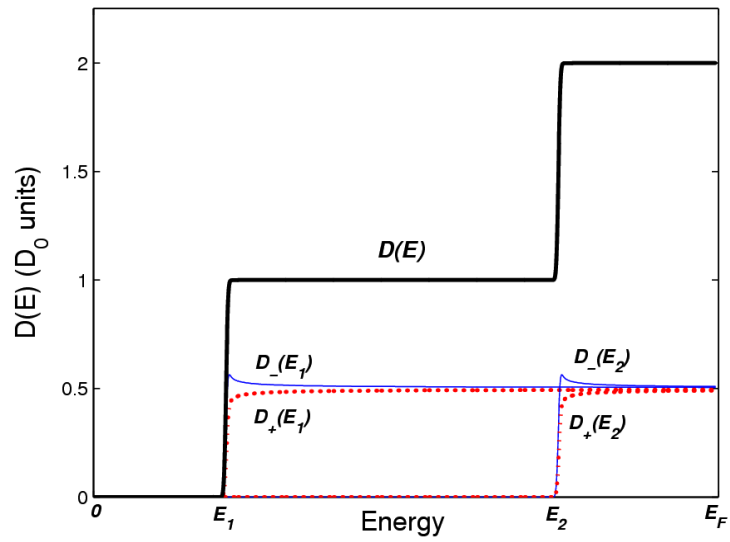

**Figure 7.** Density of states of a 2DES with Rashba SOI at zero external magnetic field. The whole DOS (black line) is the sum of the contributions of the two filled subbands with energies $E_1$ and $E_2$. Also each $E_i$ subband is split in two spin system, with DOS $D(E_i)_+$ and $D(E_i)_-$. $E_F$ is the Fermi level.

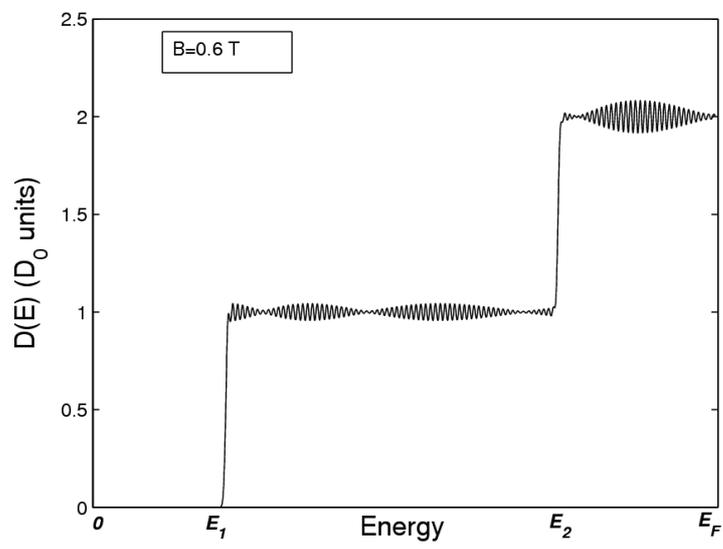

Figure 8a

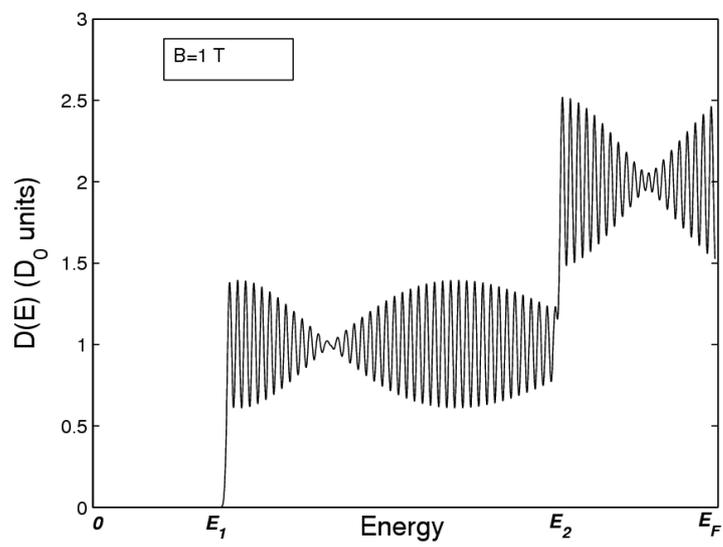

Figure 8b

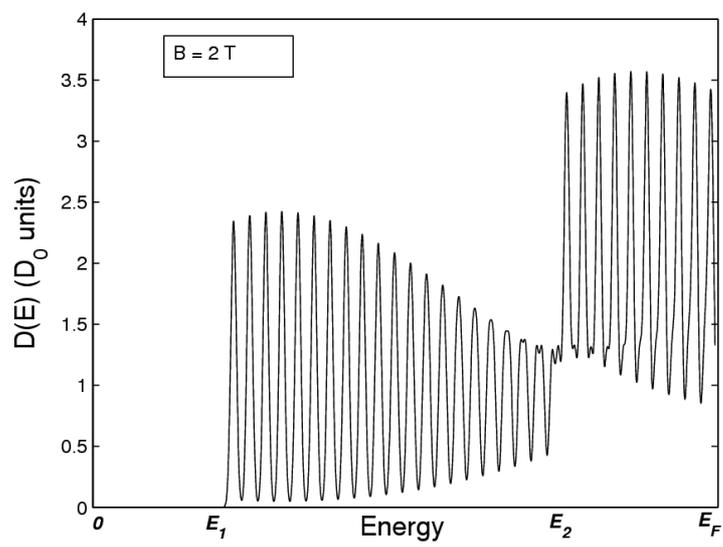

Figure 8c

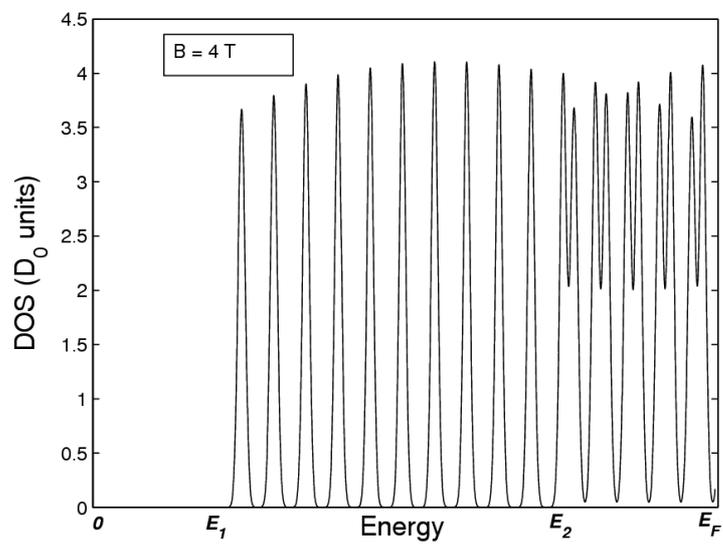

Figure 8d

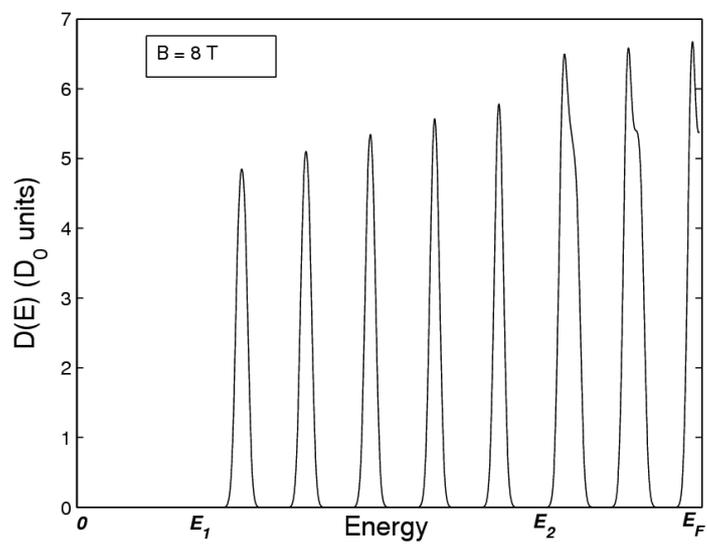

Figure 8e

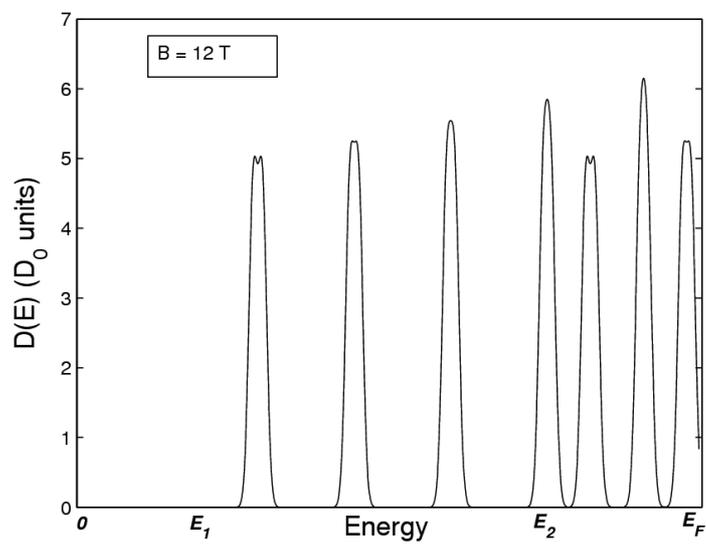

Figure 8f

**Figure 8a** to **Figure 8f** show the evolution of the density of states of a 2DES when the magnetic field increases. The electron system is confined in a QW with two filled subbands.

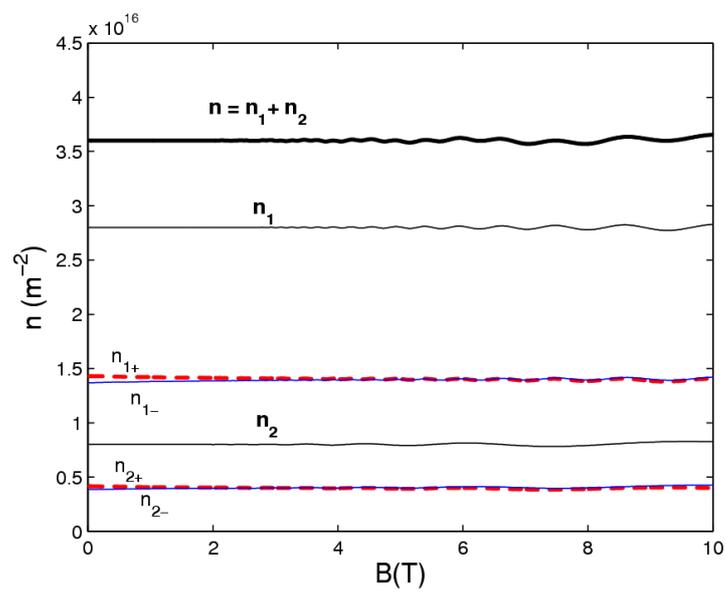

**Figure 9a** Variation of the 2DES concentration with the magnetic field. Red-dash line and blue line are related to spin-up and spin-down orientations respectively. Black thin lines are related to carrier concentration at levels $E_1$ and $E_2$, and black bold line is related to the whole 2DES carrier concentration.

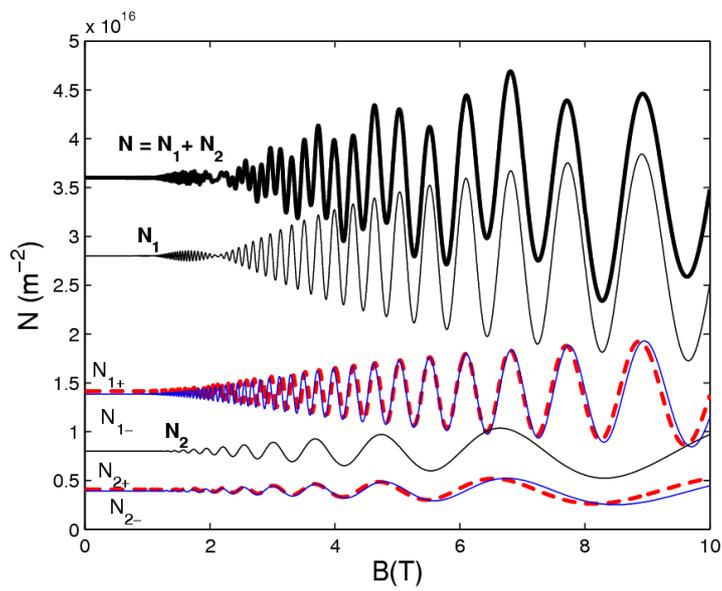

**Figure 9b** Variation of the 2DES carrier concentrations at Fermi level with the magnetic field. Red-dash line and blue line are related to spin-up and spin-down orientations respectively. Black thin lines are related to carrier concentration at levels $E_1$ and $E_2$, and black bold line is related to the whole 2DES carrier concentration.

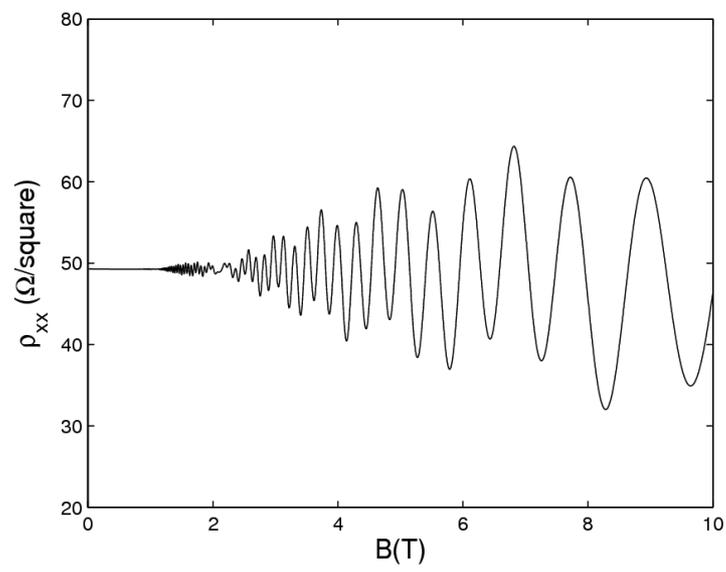

**Figure 10a** SdH oscillations beating pattern of the magnetoresistivity, with a visible node at in the region between 2 T and 2.5 T.

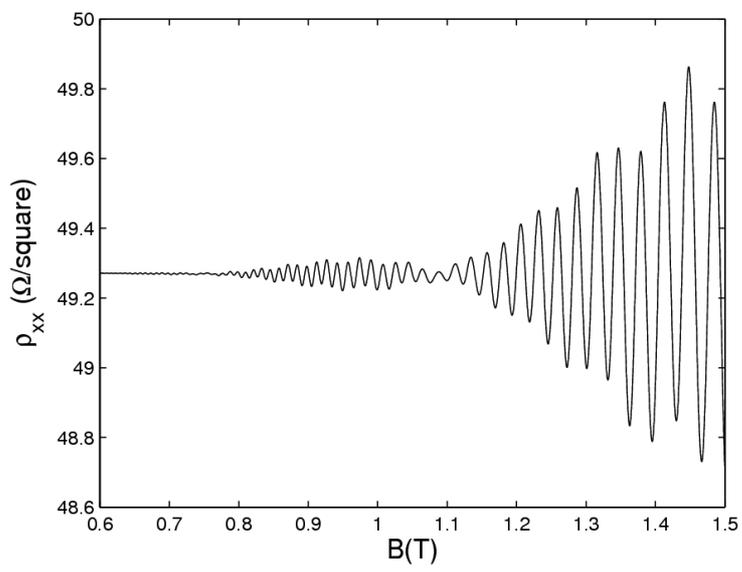

**Figure 10b** Detailed plot of the SdH oscillations that shows a clean node at 1.1 T, and other in the 0.7 T − 0.8 T interval.

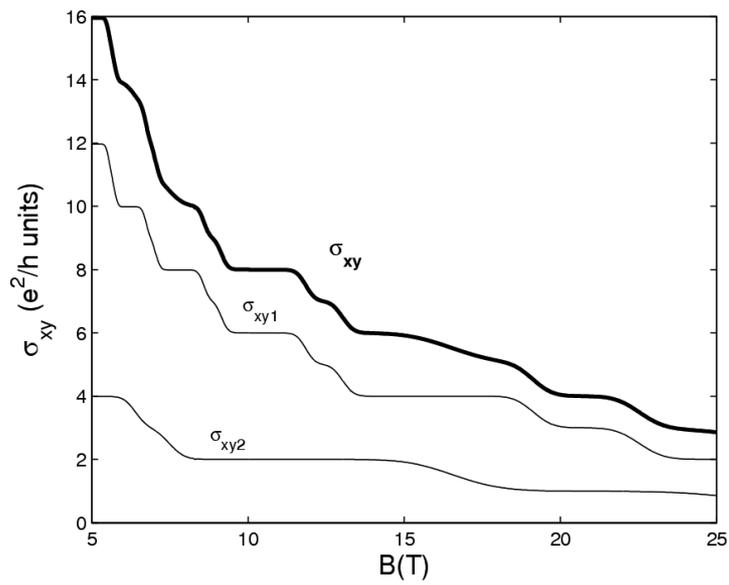

**Figure 11a** Hall Magnetoconductivity $|\sigma_{xy}|$ of the whole 2DES and magnetoconductivities $|\sigma_{xyi}|$ of the subsystems related to each filled subband vs. the external applied magnetic field.

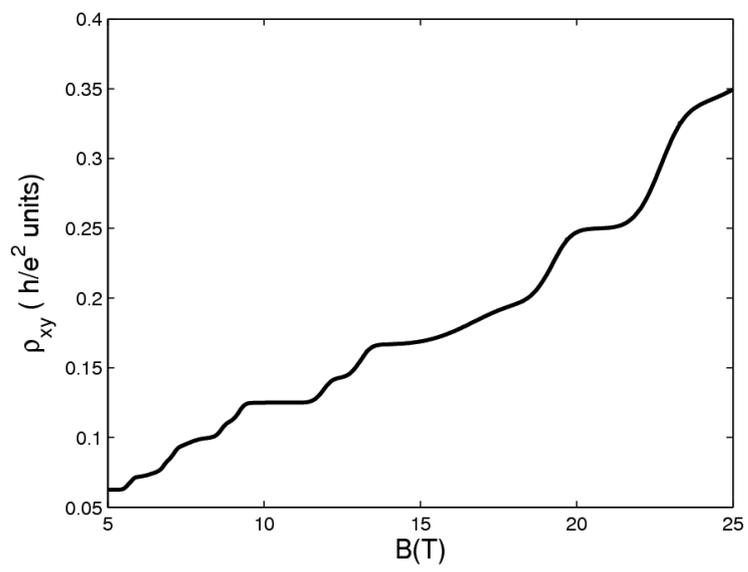

**Figure 11b** Hall Magnetoresistivity of the 2DES vs. the external applied magnetic field.